\documentstyle[12pt]{article}
\begin{document}
\thispagestyle{empty}
\begin{center}
\null
\vskip-1truecm
\rightline{IC/95/285}
\vskip1truecm International Atomic Energy Agency\\ and\\ United Nations
Educational Scientific and Cultural Organization\\
\medskip INTERNATIONAL CENTRE FOR THEORETICAL PHYSICS\\
\vskip1.5truecm {\bf
ON THE COLLECTIVE MODE SPECTRUM\\ FOR COMPOSITE FERMIONS AT 1/3 FILLING
FACTOR}\\
\vskip1.5cm
Aurora P\'erez Mart\'{\i}nez\footnote{\normalsize Permanent address: Grupo
de F\'{\i}sica Te\'orica,
Instituto de Cibern\'etica, Matem\'atica y F\'{\i}sica, Calle E No. 309 Esq. a
15, Vedado,  La Habana, Cuba.}\\
Departamento de F\'{\i}sica del CINVESTAV,\\ Apartado Postal 14-740, Mexico
07000, DF Mexico,\\
\bigskip
Alejandro Cabo$^1$\\
International Centre for Theoretical Physics, Trieste, Italy\\
\bigskip
and\\
\bigskip
Valia Guerra\\
Departamento de Matem\'atica, Instituto de Cibern\'etica,
Matem\'atica y F\'{\i}sica, \\
Calle E No.309 Esq. a 15, Vedado,
La Habana 4, Cuba.
\end{center}
\vskip0.5truecm
\centerline{ABSTRACT}
\bigskip

The collective mode spectrum of the composite fermion state at 1/3 filling 
factor is evaluated.
At zero momentum, the result coincides with the cyclotron energy at the 
external magnetic field value, and not at the effective field, in spite of the
fact that only the former  enter in the equations, thus, the Kohn theorem is
satisfied. Unexpectedly, in place of a magneto-roton minimum, the collective
mode gets a threshold indicating the  instability of the mean field composite
fermion state under the formation of crystalline  structures. However, the
question about if this outcome only appears within the mean  field approximation
should be further considered. 
\vspace{1cm}
\begin{center}
MIRAMARE -- TRIESTE\\
\medskip
September 1995
\end{center}

\newpage

\section{Introduction}
In previous papers~\cite{1}-\cite{2} we have introduced a generating functional approach 
to the composite fermion model of FQHE of Jain (~\cite{3},~\cite{4}) through considering 
the statistical interaction with a slowly varying parameter taken in the fermion 
limit~\cite{4}. In ~\cite{1} the Dyson equation were
solved in the Hartree-Fock approximation. The paper~\cite{2} was devoted to construct a 
perturbative approach based on the generating functionals and mean statistical field 
values. 

In this work we continue the analysis of the implications of the general approach in 
~\cite{2}.
Concretely the Bethe-Salpeter equation and its contribution to the linear response of the 
FQHE regimen is considered in order to discuss the elementary
excitation spectrum predicted by the model. Technically, the discussion in this paper is 
very close the one in the work~\cite{5}
in which the fractional statistics gas was investigated. Here, as the FQHE problem is 
examined the discussion includes 
the Coulomb interaction
and the presence of the external magnetic field. The intra Landau level collective mode 
spectrum was already investigated in ~\cite{6} following an approach proposed by 
Feynman. A nonvanishing energy gap associated to magneto-roton excitations was 
obtained. 

Nowadays, the general point of view is that the gap for the $\nu=1/3$ state
should be near  0.1 $e^2/r_{o}$, a
value which is compatible with recent experimental results ~\cite{7}. Therefore, taking 
into account the also widespread interest created by the composite fermion description of 
the FQHE, in this work it is intended to determine the properties for the collective mode 
spectrum in this approach. The method employed was the numerical solution of the 
Bethe-Salpeter equation for the four-points Green's function taken in the first 
approximation for the interaction kernel. Then, as the
electromagnetic response kernel is linearly given in terms of the four-point function at 
coinciding points, it follows ~\cite{5} that the collective mode spectrum 
can also be
obtained as a byproduct. 

Section 2 is devoted to review the previous introduced general background. The
scheme  of calculation is presented in Section 3. 
In Section 4 the results for the collective mode spectrum are given.

\section{Review}
In ~\cite{2} a functional way of description for the composite 
fermion model was proposed. The
treatment was based in the Green's function generating functional

\begin{eqnarray}
Z[J_{i},J_{0},\eta^*,\eta,j_{i},j_{0}] &=& \int {\cal D}a_{i} {\cal D}a_{0}{\cal D} \psi^*{\cal D} 
\psi Z[\psi,\psi^*,A_{\nu},a_{i},a_{0},\zeta ] \nonumber\\ & &. exp{ \int (-\psi^*\eta + 
\eta^*\psi + j_{i}a_{i} + j_{0}a_{0} 
+ \sigma \zeta)d^2xdx_{4}}.
\nonumber\\
& & exp{\int (J_{i}A_{i} + J_{0}A_{0}]d^3xdx_{4}}) , \end{eqnarray}

\noindent
where the arguments of $Z$ are sources for all the fields to be specified below and

\begin{equation}
\hat{Z}[\psi,\psi^*,A_{\mu},a_{i},a_{0},\zeta]=exp(S), 
\end{equation}

\noindent
in which the action in terms of the composite fermion, 
statistical and electromagnetic fields
has the form

\begin{eqnarray*}
S=\frac{1}{\hbar c}\left [\int
d^2xdx_{4}\psi^*(x)(-c\hbar\frac{\partial}{\partial x_{4}} 
-\frac{(p+e/cA^T)^2}{2m}-ieA_{4} + \mu )\psi(x)\right. \end{eqnarray*}
\[-\int\frac{e^2}{2m^*c^2}
a_{0}(x)[(\frac{e^2}{2\theta})^2(\epsilon^{ij}\partial_{i}a_{j})+ e\psi^*(x)\psi(x)]d^2xdx_4\]
\[+\int\zeta(\partial_{i}a_{i})d^2xdx_{4} + V_{I}(\psi^*,\psi)\] \begin{equation}
-\left.\frac{1}{16\pi}\int(\partial_{\mu}{\cal A}_{\nu}-\partial_{\nu}{\cal 
A}_{\mu})^2d^3xdx_{4}\right], 
\label{for}
\end{equation}
the interaction vertex $V_{I}$ is defined by 

\begin{eqnarray}
V_{I}&=&-\frac{1}{2}\int
[\psi^*(x)\psi^*(x')U({\bf x}-{\bf 
x'})\psi(x')\psi(x)]_{x_{4}=x'_{4}}d^2xd^2x'dx_{4}\nonumber\\ 
&&
+[\psi^*(x)\psi(x)U({\bf x}-{\bf x'})]_{x_{4}=x'_{4}}n_{0}d^2xd^2x'dx_{4}\nonumber\\ 
&&-\frac{e^2}{2m^*c^2}\int
d^2x'd^2xdx_{4}[\psi^*(x)\psi(x')A^2({\bf x}-{\bf 
x'})\psi(x')\psi(x)]_{x_{4}=x'_{4}}.\nonumber \end{eqnarray}
\vspace {0.5cm}

Above, $n_0$ represents the compensating charge density of the jellium, 
$\psi^*$ and $\psi$ are the composite fermion fields, $a_{\mu}$ is the statistical field, 
$A_{\mu}^T= A_{\mu}^e + a_{\mu} + {\cal A}_{\mu}$ is an auxiliary variable and 
$A^e_{\mu}$ is the vector potential of the external magnetic field in the gauge 
($\vec{A}^e= 1/2\vec{B}\times\vec{x}$, $A_{4}=0$), ${\cal A_{\mu}}$ is an external 
electromagnetic field. The statistical field is related to the fermion density through the 
usual formula 

\begin{eqnarray}
a_{i}(x) &=&\int A_{i}({\bf x}-{\bf x}')\psi^*(x')\psi(x')d^2x,\nonumber\\ 
a_{4}(x)&=&0,\nonumber
\end{eqnarray}

\noindent
where, the centered at $x^{\prime}$ solenoidal vector potential $A_{i}$ is given by 

\begin{eqnarray}
A_{i}({\bf x}-{\bf x}')=\frac{\hbar \theta c}{\pi e}\epsilon^{ij}\frac{({\bf x}-{\bf x}')_{j}}{(\mid{\bf 
x} -{\bf x}'\mid)^2},\nonumber \\ 
\epsilon^{12}=-\epsilon^{21}=1,\epsilon^{11}=\epsilon^{22}=0, \end{eqnarray}
\noindent
the fractional statistical parameter $\theta$ defines the solenoid flux in (6),
which will selected here as equal to two flux quanta and flowing in the
direction opposed to the  sense in which the magnetic field points. This
selection defines the 1/3 filling factor state  ~\cite{1}.

In references~\cite{1}-\cite{2} the Dyson equation 
in the Hartree-Fock approximation was solved and various general properties of the 
exact propagator and dielectric tensor obtained. One interesting fact following was that 
the mean field solutions are exactly characterized by a constant effective field
which correspond to the external magnetic field plus an addition 
created by the statistical field.
The selfenergy spectrum was evaluated and will be used here as a necessary piece in 
discussing the BS equation. 



\section{Scheme of Calculation}
In this section the technique for the evaluation of the collective mode spectrum is 
considered. The general formulations developed in~\cite{5} were closely
followed. From  them, it becomes clear that the collective mode spectrum should
appear as included in  the spectrum of the four-point Green function. This
occurs because, at least in the  approximation to be considered, the
electromagnetic linear response kernel is also  linearly related with the
mentioned Green function through magnitudes which are not singular in the
frequency. The approach followed  consisted in reducing the Bethe-Salpeter
equation to a matrix equation by using the  magneto-exciton two-particle
wavefunctions ~\cite{8} and then solve for the frequency  numerically by
selecting a finite number of functions in the basis. 

The integral Bethe-Salpeter equation for the four-point Green function, after introducing 
the compact notation of representing spacetime arguments by integer numbers and 
space vectors by the same numbers with arrows, can be written as

\begin{equation}
{\cal F}(1,1,2,2')={\cal F}_{o}(1,1',2,2') \end{equation}

\[
+ \int\int\int d4\,d5\,d6{\cal F}_{o}(1,1',3,4){\cal W}(3,4,5,6){\cal F}(5,6,2',2), \]

\noindent
where ${\cal F}_{o}$ is the interaction free four-point function and 
the interaction kernel in the first approximation is taken as the following functional 
derivative

\[
{\cal W}={\it i}\frac{\delta {\cal W}_{HF}(3,4)}{\delta {\it G}(5,6)}
\]
in which ${\cal W}_{HF}(1,2)$ is the Hartree-Fock inverse propagator~\cite{1}, \cite{9}. 


As mentioned above, it is possible to simplify the integral equation (7) 
introducing magneto-exciton wave functions~\cite{5},~\cite{8} which have the form

\begin{eqnarray}
\Psi_{n\alpha}^{n'}&=&\frac{(-1)^n}{L}\frac{1}{(2\pi 
2^{n+n'}n!n'!)^1/2}\left[2\frac{\partial}{\partial z_{1}^*}-\frac{1}{2}z_{1}\right] 
^n\left[2\frac{\partial}{\partial z_{2}}-\frac{1}{2}z_{2}^*\right]^{n'} \nonumber \\
& & \exp[{-(|z_{1}|^2+|z_{2}|^2+|z_{\alpha}|^2)/4}] .\exp[{(z_{1}^*z_{2}+z_{1}^*z_{\alpha}-
z_{2}z_{\alpha}^*)/2}], 
\\ z&=&x+y i.
\end{eqnarray}

These two-particle states are constructed from the one-body wave functions for electrons 
and holes in the $m$th and $n$th Landau levels. They are characterized 
by a center of mass momentum which can be written in complex form as $z_{\alpha}={\it 
i} q_{x} -q_{y}$. The main simplification introduced by these states is that the 
$z_{\alpha}$ is conserved and all the matrix elements are diagonal in this quantum 
number.

Then, the integral equation (7) can be written as a linear matrix equation in the form

\begin{equation}
{\cal F}_{nm,n',m'}(\alpha,\omega)=\delta_{nn'}\delta_{mm'}{\cal F}_{o\,nm} 
\end{equation}
\[
+\sum_{kl}{\cal F}_{o\,nm}(\omega)<nm\alpha|{\cal W}|kl\alpha>{\cal 
F}_{kl,n'm'}(\alpha,\omega). \]

The matrix elements of the free four point function ${\cal F}_o$ and the interaction 
kernel ${\cal W}$ are
given by the expressions

\begin{equation}
\left\langle
\begin{array}{ll}
m'\\
m
\end{array}
\left|{\cal F} \right|
\begin{array}{ll}
n'\\
n
\end{array}
\right\rangle
=\int\int\int\int\Psi_{m\alpha}^{*m'}(\vec{1},\vec{4}){{\cal 
F}_o}(\vec{1},\vec{4}|\vec{2},\vec{3})\Psi_{n\alpha}^{n'}(\vec{2},\vec{3}) 
d\vec{1}d\vec{2}d\vec{3}d\vec{4},
\end{equation}

\begin{equation}
\left\langle
\begin{array}{ll}
m'\\
m
\end{array}
|{\cal W}|
\begin{array}{ll}
n'\\
n
\end{array}
\right\rangle
=\int\int\int\int\Psi_{m\alpha}^{*m'}(\vec{1},\vec{4}){\cal 
W}(\vec{1},\vec{4}|\vec{2},\vec{3})\Psi_{n\alpha}^{n'}(\vec{2},\vec{3}) 
d\vec{1}d\vec{2}d\vec{3}d\vec{4},
\end{equation}
\noindent
in which the dependence on the $\alpha$ quantum number has been omitted and a 
vertical representation of the bracket indices is also used. 


The interaction kernel becomes the sum of twenty contributions arising from the anyonic 
like interactions and two from the Coulomb one in the form

\[
{\cal W}=\sum_{i=1}^{20}{\cal W}_{i}^{A} + \sum_{i=1}^{2}{\cal W}_{i}^{C}. \]

The explicit formulae for these terms and the one for ${\cal F}_{o}$ are given in Appendix 
A. 
 obtained 
The Feynman diagram representation of these contributions can be found 
in ~\cite{5}. 

Let us discuss now the parameters characterizing our specific problem. The statistical 
parameter will be selected as satisfying 
$\theta=-2k\pi$. It
corresponds with the composite fermions description in which an even number of flux 
quanta are attached to each electron~\cite{1}-\cite{2}. In addition, due to the
fact that we  are interested in analyzing the 1/3 filling factor case, only one
Landau will be filled, then $k=1$ will be fixed ~\cite{1}. This selection is related with a 
negative mean statistical field which cancels 2/3 
of the external magnetic field ~\cite{1}. All the lengths and energies here can be 
expressed in units of the magnetic length in the effective magnetic field
remaining after  the external field is partially compensated 
${\it l}_{o}$ and the
cyclotron energy	$\hbar\omega^{eff}_c$ in the same field, respectively.

The collective modes of the system, correspond with the lowest 
frequencies leading to zero
eigenvalues of the inverse kernel

\begin{equation}
{\cal F}^{-1}=({\cal F}_{o})^{-1} - {\cal W}, \end{equation}
which in addition are also singularities of the dielectric response tensor component with 
both indices being temporal ~\cite{5} . 


Due to the fact that ${\cal F}_{o}$ is qualitatively different for particle-hole and hole-
particle channels it becomes useful to divide the matricial equation (10) in four sub-blocks. 
Concretely, the block representation is obtained by restricting, 
as conceived in ~\cite{5}, the indices $m$ and $n$ to empty Landau levels and
$m'$  and $n'$ to filled ones and define the block matrices ${\cal E}$ and ${d}$
through
\[
\left\langle\begin{array}{ll}
m'\\
m
\end{array}
\left|{\cal W}\right|
\begin{array}{ll}
n'\\
n
\end{array}
\right\rangle \equiv
\left\langle\begin{array}{ll}
m'\\
m
\end{array}
\left|{\cal E}\right|
\begin{array}{ll}
n'\\
n
\end{array}
\right\rangle,
\]
and
\[
\left\langle\begin{array}{ll}
m'\\
m
\end{array}
\left|{\cal W}\right|
\begin{array}{ll}
n\\
n'
\end{array}
\right\rangle \equiv
\left\langle\begin{array}{ll}
m'\\
m
\end{array}
\left| d\right|
\begin{array}{ll}
n'\\
n
\end{array}
\right\rangle,
\]
\noindent
in which the properties of the magneto-exciton wave function also implies 

\[
\left\langle\begin{array}{ll}
m\\
m'
\end{array}
\left|{\cal W}\right|
\begin{array}{ll}
n\\
n'
\end{array}
\right\rangle \equiv
\left\langle\begin{array}{ll}
m'\\
m
\end{array}
\left|{\cal E}\right|
\begin{array}{ll}
n'\\
n
\end{array}
\right\rangle^*,
\]

\[
\left\langle\begin{array}{ll}
m\\
m'
\end{array}
\left|{\cal W}\right|
\begin{array}{ll}
n'\\
n
\end{array}
\right\rangle \equiv
\left\langle\begin{array}{ll}
m'\\
m
\end{array}
\left|d\right|
\begin{array}{ll}
n'\\
n
\end{array}
\right\rangle^*.
\]

The explicit expressions for the matrix elements of the matrices ${\cal E}$ and $d$ are 
given in Appendix B. The twenty anyonic terms associated to each of these blocks ${\cal 
E}$ and $d$ were received 
just by using
the almost identical expressions considered in ~\cite{5}. 

Now, introducing the matrix $\Delta\epsilon$ as 

\begin{equation}
\left\langle
\begin{array}{ll}
m'\\
m
\end{array}
\left|\Delta\epsilon\right|
\begin{array}{ll}
n'\\
n
\end{array}
\right\rangle
\equiv\delta_{mn}\delta_{m'n'}(\epsilon_{n}-\epsilon_{n'}) \end{equation}
in order to compact the blocks coming from the matrix representation of ${\cal F}_o$, the 
homogeneous Bethe-Salpeter equation (10) can be reduced to the matrix form ~\cite{5}.

\begin{eqnarray}
\left[\begin{array}{cc}
\hbar\omega-\Delta\epsilon-{\cal E}+{\it i}\eta & -d\\ -d^{+} & -\hbar\omega-\Delta\epsilon-
{\cal E}^{*}+{\it i}\eta \end{array}
\right]^{-1}
\end{eqnarray}

This representation was then used to determine numerically the collective mode 
dispersion relation. For this purpose a finite number of basis
functions corresponding to the  Landau levels index up to a maximum value were
retained in constructing the BS matrix  (15). The results are described in next
section.


\section{Results and discussion}

The roots of the determinant of the matrix (15) 
were found numerically after considering only the lowest Landau level indices for
the  magneto-exciton basis function up to a maximum value $n_c$. This number was
also  varied in order to check the convergence of the results. 


The lowest energy collective mode dispersion relation is shown in Fig.1 as a function of 
center of mass momentum taken along the $x_2$ axes. As noticed before, the
energies  are given in units of the cyclotron frequency of the effective
magnetic field 
$\omega_{c}^{eff}$ and the momentum
in units of $ql_{o}$. The curve corresponds to a number $n_{c}=15$ 
of Landau levels.

It can be observed that the collective mode energy decreases as the center of
mass  momenta grows up to a threshold value near $q{\it l}_{0}= 2.165$. At this
point unstable  modes having imaginary part of the frequency appear. These kind
of modes remain  up to near the $q{\it l}_{0}=3.3$
momentum value. Then, the unstability disappear and the real energy begin to grow with 
the momentum of center of mass. It can be noticed that unstability region begins for 
wavevector values greater than $q{\it l}_{0}=1$. 

The above described behaviour is not significantly affected by the presence of the 
Coulomb interaction.
The growing and spatially periodic oscillations have a wavelength of the order of the 
magnetic length in the effective magnetic field.

An important property which follows from Fig. 1 is that the Kohn theorem is exactly 
satisfied at zero momentum. That is, the collective mode energy became equal to three 
times $\hbar\omega^{eff}_{c}$, or what is the same, equal to the cyclotron
energy in the external magnetic field $\hbar\omega_{c}$. This outcome is an 
independing checking of the calculation in the low momentum limit. Another verification 
performed consisted in taking the parameter $\theta=1$ by also excluding the
Coulomb  interaction. After this, it was possible to reproduce the collective
mode dispersion for a  bose fluid obtained in Ref.~\cite{5}. Finally, previous
results for the inter Landau level  collective excitation of Kallin and Halperin
were also repeated by considering the coulomb  interaction and assuming zero
flux for the statistical solenoidal interaction, that is taking 
$\theta=k=0$. 

It is known that the perturbative calculation of the energy within the composite fermion 
approach has difficulties. One of them, for example, is that after disconnecting the 
coulomb interaction the calculated values can hardly interpreted
as reproducing the properties of the noninteracting electron system. 
Therefore, it cannot be 
discarded that the detected unstability could disappear after including correlations in the 
same way that those correlations are expected to repair the above mentioned troubles 
with energy. However, under the acceptance that the results remain valid in the
exact  theory, the unstability could be imagined to be related with the largely
discussed question  about the role of crystal fluctuations in the FQHE ground
state. Further analysis related  with these possibilities will be considered
elsewhere.

\section*{Acknowledgments}

It is pleasure to express our gratitude to our colleague Augusto Gonzalez for important 
discussions and to Profs. J.K. Jain and G. Baskaran for their comments. We also 
appreciate the suggestions made by Gerardo Gonzalez. One of the authors (A.C.)
thanks  the hospitality of the International Centre of
Theoretical Physics and the International School of Advanced Studies of Trieste during 
his stay at those institutions.
(A.P.) would like to thank the CLAFM and Department of Physics 
of CINVESTAV for their hospitality.
We are also deeply indebted to the Third World Academy of Sciences for its
support through the TWAS Research Grant 93 120 RG PHYS LA. \newpage

\eject
\noindent
{\large \bf Appendix A}

\medskip
\medskip
\noindent
{\bf a. The selfenergies and their statistical and coulomb contributions for each Landau 
level of index $n$} 

\[
{\epsilon}_{n}={\epsilon}_{H}(n)+{\epsilon}_{c}(n) \]

\[
{\epsilon}_{c}(n)= - u_c (\pi/2)^{1/2}(-1 + \sum_{l=0}^n\frac{(-1/2)^ln!(-1 + 2(l+k))!)} 
{(l!^2(n-l)!})
\]

\[
{\epsilon}_{H}(n)=n + \theta - {{{{\theta }^2}}\over 4} + 
\theta ^2 \left( -{1\over 2} + {\it E_{r}} \right) - 
\frac{\theta^2}{2}\left(1/2+ {1\over {2\,n\,\left( 1 + n \right) }} - 
\sum_{k = 1}^{n}{1\over k}\right),
\]

\noindent
where $n$ is the index of the Landau level in the effective field and the constant $u_c$ is 
given by

\[
u_c=({e^2\over \epsilon l_o})/({\hbar e B^{eff} \over m c}) \]

\medskip
\medskip
\noindent
{\bf b. The matrix elements of $ {\cal F}_o$ } 

\medskip
\medskip

$$
\left\langle
\begin{array}{ll}
m'\\
m
\end{array}
\left|{\cal F}_o \right|
\begin{array}{ll}
n'\\
n
\end{array}
\right\rangle
= \delta_{\alpha \beta} \delta_{mn} \delta_{m' n'} 
\left[
\begin{array}{cc}
(\hbar\omega-(\epsilon_{n}-\epsilon_{n'})+i \eta)^{-1} 
& n \,\,empty,\, n'\,\,occupied\\ 
(-\hbar\omega-(\epsilon_{n'}-\epsilon_{n})+i\eta)^{-1} 
& n'\,\, empty,\, n\,\, occupied \\ 0 & 
otherwise.
\end{array}
\right.
$$

\medskip
\medskip
\noindent
{\bf c. The twenty two contributions to the matrix elements of 
interaction kernel {${\cal W}$} }

\medskip

In the formulae below, the $\Psi_A$ and $\Psi_B$ represents any two states of the 
magnetoexciton basis, $\bar{A}$ is the vector potential for the effective magnetic field 
and $A_{12}$ means the solenoid vector (6) with spatial arguments 1 and 2. As in 
~\cite{3} dimensionless quantities $e,m,c$ and $\hbar$ are set equal to 1, which 
produces energies and lengths in units of $l_o$ and $\hbar w^{eff}_c$ 

\[
{\cal W}_{1}^{A}=\theta\lim_{3->1}\int d1\int d2 
\Psi_{B}(2,2)A_{12}(P_{1}+\bar{A}_{1})\Psi_{A}^*(3,1), \]

\[
{\cal W}_{2}^{A}=\theta\lim_{3->1}\int d1\int d2 
\Psi_{A}^*(2,2)A_{12}(P_{1}+\bar{A}_{1})\Psi_{B}(1,3), \]

\[
{\cal W}_{3}^{A}=-\theta\int d1\int d2 
\Psi_{B}(2,1)[A_{12}(P_{1}+\bar{A}_{1})\Psi_{A}^*(2,1), \]

\[
{\cal W}_{4}^{A}=-\theta\int d1\int d2 
\Psi_{A}^*(1,2)[A_{12}(P_{1}+\bar{A}_{1})\Psi_{B}^*(1,2), \]

\[
{\cal W}_{5}^{A}=\theta^2\int d1\int d2|A_{12}|^2\Psi_{A}^*(2,2)\Psi_{B}(1,1), \]

\[
{\cal W}_{6}^{A}=-\theta^2\int d1\int d2|A_{12}|^2\Psi_{A}^*(1,2)\Psi_{B}(1,2), \]

\[
{\cal W}_{7}^{A}=-\theta^2\int d1\int d2\int 
d3A_{12}A_{13}\Pi(1,1)\Psi_{A}^*(2,3)\Psi_{B}(2,3), \]

\[
{\cal W}_{8}^{A}=-\theta^2\int d1\int d2\int 
d3A_{12}A_{13}\Pi(1,3)\Psi_{A}^*(1,3)\Psi_{B}(2,2), \]

\[
{\cal W}_{9}^{A}=-\theta^2\int d1\int d2\int 
d3A_{12}A_{13}\Pi(3,1)\Psi_{A}^*(2,2)\Psi_{B}(1,3), \]

\[
{\cal W}_{10}^{A}=-\theta^2\int d1\int d2\int 
d3A_{12}A_{13}\Pi(1,2)\Psi_{A}^*(3,3)\Psi_{B}(2,1), \]

\[
{\cal W}_{11}^{A}=-\theta^2\int d1\int d2\int 
d3A_{12}A_{13}\Pi(2,1)\Psi_{A}^*(2,1)\Psi_{B}(3,3), \]

\[
{\cal W}_{12}^{A}=-\theta^2\int d1\int d2\int 
d3A_{12}A_{13}\Pi(3,2)\Psi_{A}^*(1,1)\Psi_{B}(2,3), \]

\[
{\cal W}_{13}^{A}=-\theta^2\int d1\int d2\int 
d3A_{12}A_{13}\Pi(2,3)\Psi_{A}^*(2,3)\Psi_{B}(1,1), \]

\[
{\cal W}_{14}^{A}=\theta^2\int d1\int d2\int 
d3A_{12}A_{13}\Pi(1,1)\Psi_{A}^*(3,3)\Psi_{B}(2,2), \]

\[
{\cal W}_{15}^{A}=\theta^2\int d1\int d2\int 
d3A_{12}A_{13}\Pi(1,3)\Psi_{A}^*(2,3)\Psi_{B}(2,1), \]

\[
{\cal W}_{16}^{A}=\theta^2\int d1\int d2\int 
d3A_{12}A_{13}\Pi(3,1)\Psi_{A}^*(2,1)\Psi_{B}(2,3), \]

\[
{\cal W}_{17}^{A}=\theta^2\int d1\int d2\int 
d3A_{12}A_{13}\Pi(1,2)\Psi_{A}^*(1,3)\Psi_{B}(2,3), \]

\[
{\cal W}_{18}^{A}=\theta^2\int d1\int d2\int 
d3A_{12}A_{13}\Pi(2,1)\Psi_{A}^*(1,3)\Psi_{B}(1,3), \]

\[
{\cal W}_{19}^{A}=\theta^2\int d1\int d2\int 
d3A_{12}A_{13}\Pi(3,2)\Psi_{A}^*(1,2)\Psi_{B}(1,3), \]

\[
{\cal W}_{20}^{A}=\theta^2\int d1\int d2\int 
d3A_{12}A_{3,2}\Pi(2,1)\Psi_{A}^*(3,1)\Psi_{B}(2,1), \]


\[
{\cal W}_{21}^{C}=-\int d1\int d2 {u_c \over |1-2|}\Psi_{A}^*(1,2)\Psi_{B}(1,2), \]

\[
{\cal W}_{22}^{C}=\int d1\int d2\int d3 {u_c \over |1-2|} \Psi_{A}^*(2,2)\Psi_{B}(1,1). \]

\noindent
where the projection operator in the first Landau level $\Pi$ is defined as

\[
\Pi(1,2)={1\over 2 \pi l_0^2 } \exp {[-(|z_1|^2+|z_2|^2)/4+z_{1}^{*} z_2/2}]. \]

As all the vectors appearing are spatial ones the arrow over them has been omitted for 
avoid more cumbersome writing. \eject
\medskip
\noindent
{\large\bf Appendix B}

\medskip
\medskip
\medskip

\noindent
{\bf a. Definitions of various auxiliary quantities} 

\[
c(m,n)=(-1)^{m + n}\frac{z_{\alpha}^n z_{\alpha}^{*m}}{(2^{m+n}m!n!)^{1/2}}, \]

\[
d(m,n)=(-1)^{m + n}\frac{z_{\alpha}^{*(m+n)}}{(2^{m+n}m!n!)^{1/2}}, \]

\[
co(m,n)=\frac{m!}{(m-l)!l!},
\]

\[
b=\frac{1}{2}|z_{\alpha}|^2
\]

\[
f_{n}=e^{-b}\sum_{k=0}^n\frac{b^k}{k!},
\]

\[
g_{n}=\sum_{k=0}^n \frac{k!}{b^{k+1}}f_{k}, \]

\[
h_{n}=\sum_{k=0}^n \frac{k!}{b^{k+1}}[1-f_{k}], \]

\[
E_{b}=-\frac{1}{2}\sum_{p=1}^{\infty}\frac{(-b)^p}{p!p}, \]

\[
X=e^{-b}\sum_{p=1}^{\infty}\frac{(b)^p}{p!p}, \]

\[
Y_{m,n}=m!n!e^{-b}\sum_{p=1}^{\infty}\frac{(p-1)!}{p+m)!(p+n)!}b^p, \]

\noindent
{\bf b. Divergent integrals }
\medskip
\medskip

\[
I_{1}=e^{-b}\lim_{\eta->0}\int_0^{\infty} dr\frac{r}{r^2+\eta^2}J_{0}(qr), \]

\[
I_{2}=e^{-b}\lim_{\eta->0}\int_0^{\infty}dr\frac{r}{r^2+\eta^2}e^{-r^2/2}, \]

\medskip
\medskip
\noindent
{\bf c. The ${\cal E}$ block matrix components for {$ m,n>0$} } 

\medskip
\medskip

\[
{\cal E}(m,n)=\sum^{22}_{r=1} {\cal E}_{r}(m,n),\\ \]

\begin{equation}
{\cal E}_{1}(m,n)=-\theta\frac{c(m,n)}{2}e^{-b}\left(\frac{m}{b}-1\right),\nonumber 
\end{equation}
\[
{\cal E}_{2}(m,n)=-\theta\frac{c(m,n)}{2}e^{-b}\left(\frac{n}{b}-1 \right), \]

\[
{\cal E}_{3}(m,n)=-\theta\frac{c(m,n)}{2}e^{-b},\\ \]

\[
{\cal E}_{4}(m,n)|_{n\geq m}=-\frac{\theta}{2}\left [ \delta_{mn} + c(m,n)\left ( 
\frac{m!}{b^m}f_{m}e^{-b} +
\frac{n!}{b^n}[1 -f_{n - 1}]\right )\right ],\\ \]

\[
{\cal E}_{5}(m,n)=\theta^2c(m,n){\it I}_{1},\\ \]

\[
{\cal E}_{6}(m,n)|_{n\geq m}=-\theta ^2\frac{c(m,n)}{2} 
\left( 2\,{\it I}_{2} + X + g_{m - 1} -
h_{n - 1} \right), \\
\]

\[
{\cal E}_{7}(m,n)|_{n\geq m}=\theta^2
\left [ {\it E}_{b} - {\it E}_{r} +\frac{e^{-b}}{2}\sum_{p = 0}^{n - 1}\frac{b^p}{p!}\left 
[\sum_{k=p+1}^{n}\right. 
\frac{1}{k}\right ]
- \frac{c(m,n)}{2} (1 - \delta_{mn})
\]

\[
.\left.\frac{1}{(n - m)}\left(\frac{m!f_{m - 1}}{b^m} + 
\frac{n!}{b^n}[ 1 - f_{n - 1}]\right)\right ], \]

\[
{ \cal E}_{8}(m,n)=\theta^2\frac{c(m,n)}{4}e^{-b}\left(\frac{m}{b}-1\right),\\ \]

\[
{\cal E}_{9}(m,n)=\theta^2\frac{c(m,n)}{4}e^{-b}\left( \frac{n}{b}-1 \right),\\ \]

\[
{\cal E}_{10}(m,n)=-\theta^2\frac{c(m,n)}{4} e^{-b}\left( \frac{1}{b}+\frac{1}{n+1} \right),\\ \]

\[
{\cal E}_{11}(m,n)=-\theta^2\frac{c(m,n)}{4}e^{-b}\left( \frac{1}{b}+\frac{1}{m+1}\right),\\ \]

\[
{\cal E}_{12}(m,n)=-\frac{\theta^2}{2}c(m,n)\left [({\it I}_{1} - {\it I}_{2}) 
+ \frac{e^{-b}}{2}\sum_{k = 1}^{n}\frac{1}{k}\right ], \\ \]

\[
{\cal E}_{13}(m,n)=-\frac{\theta^2}{2}c(m,n)\left [({\it I}_{1} - {\it I}_{2}) 
+ \frac{e^{-b}}{2} \sum_{k = 1}^{m}\frac{1}{k}\right ], \]

\[
{\cal E}_{14}(m,n)=\theta^2\frac{c(m,n)}{2}e^{-b}\frac{1}{b},\\ \]

\[
{\cal E}_{15}(m,n)=\theta^2\frac{c(m,n)}{4}e^{-b},\\ \]

\[
{\cal E}_{16}(m,n)=\theta^2\frac{c(m,n)}{4}e^{-b}, \]

\[
{\cal E}_{17}(m,n)|_{n\geq m}=\theta^2 \frac{c(m,n)}{4}\left( \frac{1}{n}\frac{n!}{b^n}[1 - 
\delta_{mn} - 
f_{n - 1}] +\frac{1}{n+1}\frac{m!}{b^m}f_{m} \right), \]

\[
{\cal E}_{18}(m,n)|_{n\geq m}=-\frac{\theta^2}{4}c(m,n)\left [\frac{1}{m}\frac{m!}{b^m}f_{m-
1} + \frac{1}{m+1}\frac{n!}{b^n}[(1 - \delta_{mn}) - f_{n}]\right ],\\ \]

\[
{\cal E}_{19}(m,n)|_{n\geq m}=\frac{\theta^2}{4}c(m,n)\left( X+g_{m-1}-h_{n-1}+ 
\frac{m!n!}{b^{m+n+1}}f_{m}(1-f_{n})e^{b}\right), \]

\[
{\cal E}_{20}(m,n)=\frac{\theta^2}{4}c(m,n)\left(\frac{1}{b}e^{-b}+Y_{m,n}\right), \]

\[
{\cal E}_{21}(m,n)=
- u_c \,c(m,n)e^{-b}\sum_{l=0}^{m}\sum_{k=0}^n co(m,l)co(n,k)\Gamma(1+k+l+|l-
k|/2)(1/2)^{|l-k|} \]

\[
.(-1)^{l+k} |z_{\alpha}|^{l-k}
\frac{{}_{1}F_{1}(1+k+l+|l-k|/2,1+|k-l|,|z_{\alpha}|^2/2)}{{2}^{1/2}\Gamma(1+|l-k|)} \]

\[
{\cal E}_{22}(m,n)=u_c\, c(m,n)\frac{e^{-b}}{|z_{\alpha}|^2} \]

\medskip
\medskip
\noindent
{\bf d. The d block matrix components for {$ m,n>0$}} \medskip
\medskip

\[
d(m,n)= \sum^{22}_{r=1}d_{r}(m,n),
\]

\[
d_{1}(m,n)=-\theta\frac{d(m,n)}{2}e^{-b}\left(\frac{m}{b}-1\right), \]

\[
d_{2}(m,n)=-\theta\frac{d(m,n)}{2}e^{-b}\left(\frac{n}{b}-1\right), \]

\[
d_{3}(m,n)=-\theta \frac{d(m,n)}{2}
\left(e^{-b} - m\frac{(m+n-1)!}{b^{m+n}}[1-f_{n+m-1}]\right), \]

\[
d_{4}(m,n)=-\theta\frac{d(m,n)}{2}
\left(e^{-b} - n\frac{(m+n-1)!}{b^{m+n}}[1-f_{n+m-1}]\right), \]

\[
d_{5}(m,n)=\theta^2d(m,n) {\it I}_{1},
\]

\[
d_{6}(m,n)=- \theta^2\frac{d(m,n)}{2}\left(2{\it I}_{2} + X-h_{m+n-1}\right), \]

\[
d_{7}(m,n)=-\theta ^2 \frac{d(m,n)}{2}\left(\frac{(m + n-1)!}{b^{m+n}}[1-f_{m+n-1}]\right), \]

\[
d_{8}(m,n)=- \theta ^2\frac{d(m,n)}{4}e^{-b}\left(\frac{1}{m+1} + 
\frac{1}{b}\right),
\]

\[
d_{9}(m,n)= \theta ^2\frac{d(m,n)}{4}e^{-b}\left(\frac{n}{b}-1\right), \]

\[
d_{10}(m,n)=-\theta ^2 \frac{d(m,n)}{4}e^{-b}\left(\frac{1}{b}+\frac{1}{n+1}\right),	\\
\]

\[
d_{11}(m,n)=\theta^2\frac{d(m,n)}{4}e^{-b}\left(\frac{m}{b}-1\right), \]

\[
d_{12}(m,n)=-\theta ^2 \frac{d(m,n)}{2}\left ( 
({\it I}_{1} - {\it I}_{2}) + \frac{1}{2} 
{e^{-b}\sum_{k = 1}^{n}{1\over k}}\right), \]

\[
d_{13}(m,n)=\theta ^2\frac{d(m,n)}{2}\left ( 
({\it I}_{1} - {\it I}_{2}) + \frac{1}{2} 
{e^{-b}\sum_{k = 1}^{m}{1\over k}}\right), \]

\[
d_{14}(m,n)=\theta ^2\frac{d(m,n)}{2}
\frac{e^{-b}}{b},\\
\]

\[
d_{15}(m,n)=\theta ^2\frac{d(m,n)}{4}\left( 
\frac{1}{m+1}e^{-b} +\frac{( m+n-1)!}{b^{m+n}} [1-f_{m+n-1}]\right),
\]

\[
d_{16}(m,n)=\theta ^2\frac{d(m,n)}{4}\left(e^{-b}-m\frac{( m+n-1)!}{b^{m+n}} 
[1-f_{m+n-1}]\right),
\]

\[
d_{17}(m,n)=\theta ^2\frac{d(m,n)}{4}\left(\frac{1}{n+1}e^{-b}+\frac{( m+n-1)!}{b^{m+n}} + 
[1-f_{m+n-1}]\right),
\]

\[
d_{18}(m,n)=\theta ^2\frac{d(m,n)}{4}\left(e^{-b}-n\frac{( m+n-1)!}{b^{m+n}} 
[1-f_{m+n-1}]\right),
\]

\[
d_{19}(m,n)=\theta ^2\frac{d(m,n)}{4}\left(X-h_{m+n-1}+e^{-
b}\sum_{k=1}^m\frac{1}{k}\right), \]

\[
d_{20}(m,n)=\theta ^2\frac{d(m,n)}{4}\left( X-h_{m+n-1}+e^{-
b}\sum_{k=1}^n\frac{1}{k}\right), \]

\[
d_{21}(m,n)=-u_c\, d(m,n)e^{-
b}\sum_{l=0}^{m+n}co(m,l)(1/4)^{l/2}|z_{\alpha}|^{l/2}\Gamma(l+1/2) \]

\[
.\frac{{}_{1} F_{1}(l+1/2,1+l+1,|z_{\alpha}^2/2)}{\sqrt{2}(1/2)^l\Gamma(l+1)}, \]

\[
d_{22}(m,n)=u_c \, d(m,n)\frac{e^{-b}}{|z_{\alpha}|}. \]



\eject

\noindent
{\large \bf Figure Caption}

\medskip
\medskip

Fig. 1 The collective mode energy in units of $\hbar w^{eff}_c$ as a function 
of the center of mass momentum. Note that the Kohn theorem is satisfied at zero 
momentum and that the system develops a threshold after which 
unstable modes appear in a band of momentum values.

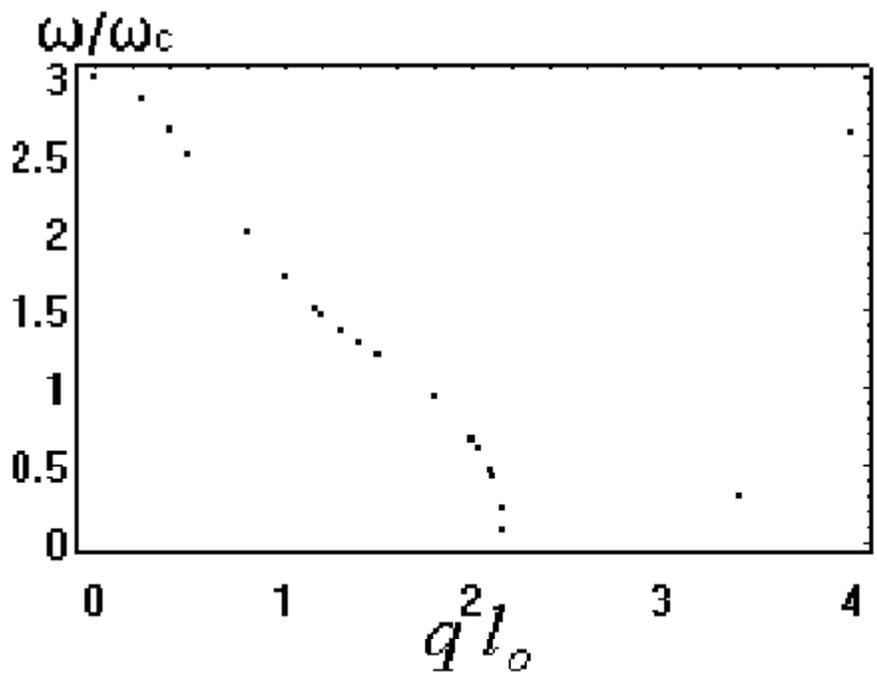
\end{document}